\documentclass[a4paper,reqno,12pt,draft]{article}
\usepackage{amssymb,euscript}

\usepackage{epsfig}

\newcommand{\M}{{\cal M}}

\newcommand{\nn}{\nonumber\\}
\newcommand{\p}{\partial}
\newcommand{\ie}{{\it i.e.}}

\newcommand{\be}{\begin{equation}}
\newcommand{\ee}{\end{equation}}
\newcommand{\ba}{\begin{eqnarray}}
\newcommand{\ea}{\end{eqnarray}}

\newcommand{\ft}{\footnote}

\newcommand{\Tr}{{\rm Tr}}

\begin{document}

\begin{flushright}
\end{flushright}
\begin{flushright}
\end{flushright}
\begin{center}
\Large{\sc Distinguishing Off-Shell Supergravities With On-Shell Physics}\\
\bigskip
\large {\sc Neil D. Lambert
\ft{lambert@mth.kcl.ac.uk}\\
\smallskip\large
{\sf Dept. of Mathematics\\
King's College London\\
The Strand\\
London, UK\\ 
WC2R 2LS\\}}
\medskip
and\\
\medskip
\large {\sc Gregory W. Moore
\ft{gmoore@physics.rutgers.edu}\\
\smallskip\large
{\sf Dept. of Physics and Astronomy\\
Rutgers University\\
136 Frelinghuysen Road\\
Piscataway, NJ\\ 
08854, USA\\}}
\end{center}

\bigskip
\begin{center}
{\bf {\sc Abstract}}
\end{center}

We show that it is possible to distinguish between 
different off-shell completions of supergravity at the on-shell level.
We focus on the comparison of the   ``new minimal'' formulation of off-shell 
four-dimensional $N=1$ supergravity  with  the  ``old minimal''  
formulation.   We show that 
there are   3-manifolds which admit supersymmetric 
compactifications in  the new-minimal formulation but which 
do not admit supersymmetric compactifications in other formulations. 
Moreover, on manifolds with boundary the new-minimal formulation 
admits ``singleton modes'' which are absent in other formulations.

\newpage

\bigskip

\section{\sl Introduction}

The equations of motion of four-dimensional $N=1$ supergravity can be
obtained using the Euler-Lagrange equations applied to the Lagrangian
\footnote{ Here $\mu,\nu,...=0,1,2,3$ are world indices and an underline 
denotes the tangent frame. We use the $(-,+,+,+)$ signature with 
$\{\Gamma_{\mu},\Gamma_{\nu}\} = 2g_{\mu\nu}$, 
$\Gamma_{5}={1\over 24}\epsilon^{\mu\nu\lambda\rho}
\Gamma_{\mu\nu\lambda\rho}$ and $\epsilon^{0123}=1$.}
\be
{\cal L}_{on-shell} = eR - 4i\bar\psi_\mu\Gamma^{\mu\nu\rho}{D}_\nu
\psi_{\rho}
\ee
While this Lagrangian is invariant, up to a total derivative,  under the transformation
\be
\delta_\epsilon e_\mu^{\ \ \underline \nu} 
= -2i\bar\epsilon\Gamma^{\underline \nu}\psi_{\mu}\ ,\qquad
\delta_\epsilon\psi_\mu = {D}_{\mu}\epsilon 
\ee
these transformations 
do  not close to the supersymmetry algebra unless the fields are
taken to be on-shell. Indeed off-shell there are only six Bosonic degrees
of freedom whereas there are twelve Fermionic degrees of freedom.
Thus it is of interest to construct off-shell extensions of
the supergravity Lagrangian.

The original motivations for studying off-shell completions 
of supergravity  were to ensure that
supersymmetry remains a valid symmetry at the quantum level as well as to 
facilitate the proof of non-renormalization theorems. 
A natural starting point for
supergravity is the geometrical analysis of four-dimensional $N=1$
superspace. Four-dimensional $N=1$ superfields carry
reducible  supersymmetry multiplets. Therefore additional constraints need
to be imposed to truncate the superfields. 
These are then combined with  the torsion and Bianchi identities
to solve for the independent fields. 
 It turns out that  
there are various  ways to do this and hence there are several different 
off-shell formulations. The predominant view of these various off-shell
completions   is that they are all equivalent on-shell.
The purpose of this paper is to show that while this is locally true, 
it is globally false.

In this paper we will explore some novel aspects of  an off-shell
formulation of  four-dimensional $N=1$ supergravity known as  ``new
minimal supergravity'' (NMS) \cite{SW}.   We show that NMS admits  Killing spinors on
manifolds which are not supersymmetric in other formulations such as 
the   ``old-minimal'' formulation 
\cite{Ferrara:1978em,Stelle:1978ye,Kaku:1978ea,BaggerWess,West:1990tg}. Moreover,  
when formulated on a manifold with boundary certain gauge modes of the 
 auxiliary fields of NMS can become dynamical (depending on boundary conditions).

From a string theory perspective off-shell formulations are often 
viewed as   unnecessary  luxuries since one is
simply viewing supergravity as a low energy effective theory which
reproduces the correct on-shell physics. Our results call that point 
of view into question.  A key motivation for this study was the 
desire to formulate 
11-dimensional supergravity on Spin$^c$ manifolds
(see the discussion section below). 
Indeed NMS contains an auxiliary gauge field which allows one to 
define the theory (off-shell) on Spin$^c$ manifolds.  
Unfortunately very little is known about off-shell completions of 
11-dimensional supergravity. Indeed it is generally
believed that an infinite number of auxiliary  fields are required
(although in some circumstances one could consider a finite collection
of auxiliary fields which do not entirely close the algebra 
\cite{deWit:1981ug}).  Our results raise the important question of 
whether different off-shell formulations of M-theory could  be 
physically inequivalent.

\section{\sl Old and New Minimal Supergravity}

The most familiar off-shell completion of $N=1$ four-dimensional supergravity
is the so-called old-minimal formulation which
includes two real scalar fields $M$ and $N$ along with a one-form
$b$ \cite{Ferrara:1978em,Stelle:1978ye,Kaku:1978ea,BaggerWess,West:1990tg}. The action is simply
\be
{\cal L}_{old-minimal} = {\cal L}_{on-shell} -\frac{1}{2}eM^2
-\frac{1}{2}eN^2 +\frac{1}{2} b\wedge\star b
\ee
and there are twelve Bosonic and twelve Fermionic degrees of freedom off-shell.
Clearly these new fields do not alter the theory in any non-trivial
way. However one does find that the supersymmetry algebra closes
off-shell (along with appropriate modifications to the supertransformation  
rules to include the auxiliary fields).

In \cite{SW} Sohnius and West gave an off-shell formulation of
four-dimensional $N=1$ supergravity which, in addition to the
graviton and gravitini, includes an auxiliary 1-form $A=A_\mu dx^\mu$ 
and a 2-form 
$B={1\over2}B_{\mu\nu}dx^\mu\wedge dx^\nu$. The Lagrangian is
\be
{\cal L}_{NMS} = eR - 4ie\bar\psi_\mu\Gamma^{\mu\nu\rho}{\cal D}^+_\nu
\psi_{\rho} - {6}V\wedge\star V - 4A\wedge dB
\label{lagrangian}
\ee
where
\begin{eqnarray}
V &=& -{1\over 2}\star \left(dB - {i} \bar \psi_\nu\Gamma_{\lambda}
\psi_{\rho}dx^\nu\wedge
dx^\lambda\wedge dx^\rho\right)
\nonumber\\
{\cal D}^+_{\mu}\psi_\nu &=& D_{\mu}\psi_\nu + A_\mu \Gamma_5\psi_\nu
\nonumber\\
\omega_{\mu\underline {\lambda\rho}}
&=& \omega^{{\rm Levi-Civita}}_{\mu\underline {\lambda\rho}}
-i(\bar\psi_{\underline \lambda}\Gamma_{\mu}\psi_{\underline \rho}
+\bar\psi_{\mu}\Gamma_{\underline \lambda}\psi_{\underline \rho}
-\bar\psi_{\mu}\Gamma_{\underline \rho}\psi_{\underline \lambda})
\end{eqnarray}

The equation of motion for $A$ sets 
\be
V=0
\ee
and then the equation of motion for $B$ determines that $A$ is
a flat connection
\be
dA=0
\ee
The particular choice $A=0$ then leads to the usual equations of motion
for the graviton and gravitini.

In addition to diffeomorphisms the theory is invariant 
under the local supersymmetry transformation 
\begin{eqnarray}
\delta_\epsilon e_\mu^{\ \ \underline \nu} 
&=& -2i\bar\epsilon\Gamma^{\underline \nu}\psi_{\mu}\nonumber\\
\delta_\epsilon\psi_\mu &=& {\cal D}^+_{\mu}\epsilon  - V_{\mu}\Gamma_5\epsilon
+{1\over 2}\Gamma_{\mu}^{\ \ \nu}\Gamma_5V_\nu\epsilon \nonumber\\
\delta_\epsilon B_{\mu\nu} &=& 4i\bar\epsilon \Gamma_{[\mu}\psi_{\nu]}
\nonumber\\
\delta_\epsilon A_{\mu}&=& -2i\bar\epsilon\Gamma_5\Gamma_{\mu}\Gamma^{\nu\lambda}\left( {\cal D}^+_\nu\psi_\lambda  + 3\Gamma_5\psi_\nu V_\lambda 
+{3\over 2}\Gamma_{\nu}^{\ \ \rho}\psi_\lambda V_\rho \right)\nonumber\\
\end{eqnarray}
However the crucial difference \cite{deWit:1981fh} between old and new minimal supergravity is
that NMS is also invariant  
under  a local chiral rotation
\begin{eqnarray}
\delta_\chi e_\mu^{\ \ \underline \nu} 
&=& 0\nonumber\\
\delta_\chi\psi_\mu &=& -\chi\Gamma_5\psi_{\mu}\nonumber\\
\delta_\chi B &=& 0
\nonumber\\
\delta_\chi A&=&d\chi. \nonumber\\
\end{eqnarray}
which is broken in old minimal by the supersymmetry
transformation rules. 
The Lagrangian also has a trivial gauge transformation which only acts 
on $B$
\be
\delta_{\lambda}B = d\lambda.
\ee
It is easy to check that all these symmetries commute with each other
\be
[\delta_\epsilon,\delta_\chi]=[\delta_\epsilon,\delta_\lambda]=[\delta_\chi,\delta_\lambda]=0
\ee
provided that the supersymmetry generator is also taken to transform
under local chiral rotations. 

Under supersymmetry
the action  is not invariant but rather transforms into a boundary term.
To quadratic order in the Fermions one finds
\begin{eqnarray}
\delta_\epsilon S &=&\int_\M \delta_\epsilon {\cal L}_{NMS}\nn
&=& -2i\int_{\p\M}\sqrt{-h}
(h^{\rho\nu}n^\mu+h^{\rho\mu}n^\nu-2h^{\mu\nu}n^\rho)D_\rho(\bar\epsilon
\Gamma_\mu\psi_\nu)
\nn&&+
4i\int_{\p\M}\bar\epsilon \left( \Gamma_5\Gamma_\lambda{\cal D}^-_{\nu}\psi_\rho +\Gamma_\lambda V_\nu\psi_\rho
+{1\over 2}\Gamma_{\nu}^{\ \
  \tau}\Gamma_\lambda V_\tau\psi_\rho
\right)  dx^\lambda\wedge dx^\nu\wedge dx^\rho\nonumber\\
\end{eqnarray}
where  $n^\mu$ is the unit inward pointing normal vector to the  boundary, $
h_{\mu\nu} = g_{\mu\nu}-n_\mu n_\nu$ is the induced metric and
\be
{\cal D}^-_{\nu}\psi_\rho=
D_\nu\psi_\rho- \Gamma_5A_\nu\psi_\rho
\ee
is the anti-chiral covariant derivative, {\it i.e.} it corresponds to 
gauging chiral rotations with the opposite choice of  $\Gamma_5$.
The fact that $\delta_\epsilon{\cal L}$ is no longer chirally invariant
seems odd but can be verified by noting that, due to the final Chern-Simons
term,  the Lagranian is  not exactly chirally invariant either
\be
\delta_\chi {\cal L}_{NMS} = -4d\chi\wedge dB = -4d(\chi dB)
\ee
The failure of chiral symmetry in $\delta_\epsilon {\cal L}$ is then
necessary to account for the variation of  $\delta_\chi {\cal L}$ 
under supersymmetry since it must be true that
\be
\delta_\chi\delta_\epsilon{\cal L}_{NMS}= \delta_\epsilon\delta_\chi{\cal L}_{NMS}
\ee
which one can readily check is indeed the case with these boundary terms. 

However we can correct for this by adding the total derivative term
\be
{\cal L}_{bdry}= 4 d \left(A\wedge B \right)
\ee
to the Lagrangian. This has the effect of changing the Lagrangian to
\be
{\cal L}'_{NMS} = eR - 4ie\bar\psi_\mu\Gamma^{\mu\nu\rho}{\cal D}^+_\nu
\psi_{\rho} - 6V \wedge\star V -4dA\wedge B
\label{improvedlagrangian}
\ee
Clearly ${\cal L}'_{NMS}$ is invariant under chiral rotations, with no
boundary terms. 
This implies that the boundary terms  must be invariant under chiral rotations.
The variation of ${\cal L}_{bdry}$ under supersymmetry is
\begin{eqnarray}
\delta_{\epsilon}{\cal L}_{bdry}&=&
4d \left(\delta_{\epsilon} A\wedge B 
- 2i\bar\epsilon A_\lambda \Gamma_\nu\psi_\rho 
dx^\lambda\wedge dx^\nu\wedge dx^\rho \right)\nonumber\\
\end{eqnarray}
The second term converts the ${\cal D}^-_\mu$  into
a ${\cal D}^+_\mu$ derivative. Thus the variation of the improved action 
under supersymmetry is
\begin{eqnarray}
\delta_\epsilon S' &=&\int_\M \delta_\epsilon {\cal L}_{NMS}
+\int_{\p\M}\delta_{\epsilon}{\cal L}_{bdry}\nn
&=& -2i\int_{\p\M}\sqrt{-h}
(h^{\rho\nu}n^\mu+h^{\rho\mu}n^\nu-2h^{\mu\nu}n^\rho)D_\rho(\bar\epsilon\Gamma_\mu\psi_\nu)
\nn&& 
+4i\int_{\p\M}\bar \epsilon \left(\Gamma_5\Gamma_\lambda{\cal D}^+_\nu\psi_\rho 
+\Gamma_{\lambda}V_\nu\psi_\rho  
+{1\over2}\Gamma_{\nu}^{\ \ \tau}\Gamma_{\lambda}V_\tau\psi_\rho\right)
 dx^\nu\wedge dx^\rho\wedge dx^\lambda \nonumber\\
&&+4\int_{\p\M} \delta_{\epsilon} A\wedge B 
\nonumber\\
\label{newsuperJ}
\end{eqnarray}
which is indeed invariant under chiral rotations.

\subsection{Supersymmetry in the presence of boundaries}

It is well known that if supergravity is placed on a manifold with a 
boundary then at least half of the supersymmetries will be broken.  If $\p\M \ne 0$ then the action obtained from $\cal L'$ is not invariant
under supersymmetry. However by adding suitable boundary terms this
can be corrected. These boundary terms are used to set-up a well-posed
boundary value problem  and also to preserve half of the supersymmetries.
In the case of eleven-dimensional supergravity this has
been done in \cite{Moss} and we wish to follow a similar analysis for NMS, 
although we will restrict our attention to quadratic terms in the Fermions.

The first step to including boundaries is to  add the 
Gibbons-Hawking term to make the pure gravitational variational problem 
well posed
\be
S_{GH} =  2\int_{\p \M} \sqrt{-h}K
\ee
where 
$K_{\mu\nu} = h_{\mu\lambda}h_{\nu\rho}D^\lambda n^\rho$ is the extrinsic
curvature. With this term in
place one finds that the variation of the standard Einstein-Hilbert plus
Gibbons-Hawking term results in the boundary term
\be
\delta_{g}(S_{EH}+S_{GH})  = -\int_{\p\M}\sqrt{-h}(K_{\mu\nu}-
g_{\mu\nu}K)\delta g^{\mu\nu}
\ee
which is required to cancel with any additional stress-energy tensor that is
localized to the boundary (which in our case vanishes).

Following \cite{Moss} we need to add a Fermionic boundary term.
Let us first recall the case of the familiar on-shell supergravity. Varying the
Fermionic term gives the equations of motion plus the boundary term
\be
\delta_\psi S
= 4i\int_{\p\M}\sqrt{-h} \bar\psi_{\mu'} \Gamma^{\mu'\nu'}\Gamma^n\delta\psi_{\nu'}
\label{deltapsiS}
\ee
where $\Gamma^n = n^\mu\Gamma_\mu$ and $\mu',\nu'$ are the coordinates
tangential to the boundary.
To cancel this one adds the term \cite{LM}
\be
S_{LM} 
= 2i\xi\int_{\p \M}\sqrt{-h}\bar \psi_{\mu'}\Gamma^{\mu'\nu'}\psi_{\nu'}
\label{oldbt}
\ee 
where $\xi=\pm 1$. Variation of this term gives
\be
\delta_\psi  S_{LM} = 4i\xi\int_{\p \M}\sqrt{-h}\bar \psi_{\mu'}\Gamma^{\mu'\nu'}\delta \psi_{\nu'}
\ee
Thus a suitable boundary condition is $\psi_{\mu'} = \xi\Gamma^n\psi_{\mu'}$.

In NMS we again encounter the boundary term
(\ref{deltapsiS}) however we cannot
simply add (\ref{oldbt}) since this term is not invariant under chiral
rotations. More properly  it doesn't even make sense as $\psi_\mu$ is a 
section of a chiral spinor bundle 
whereas $C\Gamma^{\mu'\nu'}$
is a map between the chiral and  anti-chiral spinor bundles. 

If we choose a boundary condition where $A=d\Phi_A$ on $\p\M$ then we
can add the boundary term
\be
S_{\p\psi} = 2i\xi\int_{\p \M}\sqrt{-h}\bar \psi_{\mu'}\Gamma^{\mu'\nu'} e^{2\Phi_A \Gamma_5}\psi_{\nu'}
\ee
\ie\ we can map NMS to on-shell supergravity by a chiral gauge
rotation.  
(Note that this boundary condition implies that the gauge field $A$ is
trivial on the boundary and this is not the case in general.)
We now find the  Fermionic boundary condition
\be
\Gamma^n e^{\Phi_A \Gamma_5}\psi_{\nu'} =
\xi e^{\Phi_A \Gamma_5}\psi_{\nu'}
\label{Fbc}
\ee

Our next task is to show that 
\be
S =  S_{GH} + S_{\p \psi}+ \int_\M {\cal L}' 
\ee
is indeed supersymmetric with a well posed variational problem when we impose the
boundary conditions
\be
\Gamma^n e^{2\Phi_A \Gamma_5}\psi_{\nu'} =
\xi\psi_{\nu'}\ ,\qquad A = d\Phi_A, \qquad V=0
\ee
on $\p\M$.
We derived the boundary terms by ensuring a well-posed boundary value problem
for the  metric and  $\psi_\mu$. 
Thus it remains 
to check that variations of the form $\delta A = d\delta \Phi_A$ 
on the boundary are well posed, \ie\ that they do not over constrain the 
system. There are two sources for these variations,
boundary terms from the bulk $ d\delta A\wedge B$ term and also terms 
that arise directly from varying $S_{\p\psi}$. Putting these together we
find
\ba
\delta_{\Phi_A} S &=& 4\int_{\p\M}d\delta \Phi_A\wedge B  + 4i\xi
\int_{\p\M} \sqrt{-h}\delta\Phi_A \bar \psi_{\mu'}\Gamma^{\mu'\nu'}\Gamma_5 e^{2\Phi_A\Gamma_5} \psi_{\nu'} 
\nn
&=&-4\int_{\p\M} \delta \Phi_A\wedge dB  - 
i\sqrt{-h} \delta\Phi_A \bar \psi_{\mu'}\Gamma^{\mu'\nu'}\Gamma_5\Gamma^n \psi_{\nu'}
\nn
&=& -4\int_{\p\M} \delta \Phi_A\wedge * V \nn
&=&0\nn
\ea

Next we examine the variation of the action under supersymmetry. We already
know that the bulk is supersymmetric. We would like to show that the
variation of the additional boundary terms $S_{GH}+S_{\p \psi}$ cancels
(\ref{newsuperJ}). Of course the boundary condition on the Fermions
breaks half of the supersymmetries, leaving only those with 
$\Gamma^n e^{\Phi_A\Gamma_5}\epsilon=\xi e^{\Phi_A\Gamma_5}\epsilon$.
Therefore we compute 
\ba
\delta_\epsilon (S_{EH}+S_{GH})&=& 4i\int_{\p\M}\sqrt{-h}
\bar\epsilon(K^{\mu'\nu'}-h^{\mu'\nu'}K)\Gamma_{\mu'}\psi_{\nu'}\nn
\ea
and (using the boundary conditions)
\ba
\delta_\epsilon S_{\p\psi}&=& 4i\xi \int_{\p\M}\sqrt{-h}\left(
 {{\cal D}^+_{\mu'}{\bar \epsilon}} \Gamma^{\mu'\nu'}e^{2\Phi_A\Gamma_5}
\psi_{\nu'}
+ \delta_\epsilon \Phi_A
\bar \psi_{\mu'}\Gamma^{\mu'\nu'}e^{2\Phi_A\Gamma_5}\Gamma_5\psi_{\nu'}\right) \nn
&=& 4i\int_{\p\M}\sqrt{-h}\left(
{{\cal D}^+_{\mu'}{\bar\epsilon}} \Gamma^{\mu'\nu'}\Gamma^n\psi_{\nu'}
+ \delta_\epsilon \Phi_A
\bar \psi_{\mu'}\Gamma^{\mu'\nu'}\Gamma_5\Gamma^n\psi_{\nu'}\right)\nn
&=& -4i\int_{\p\M}\sqrt{-h}\left(
\bar \epsilon \Gamma^{\mu'\nu'}\Gamma^n{\cal D}^+_{\mu'}\psi_{\nu'}
+\bar \epsilon \Gamma^{\mu'\nu'} D_{\mu'}n_\lambda \Gamma^\lambda\psi_{\nu'}
\right)\nn&&
+4\int_{\p\M} \delta_\epsilon \Phi_A \wedge dB\nn
&=&-4i\int_{\p\M}\sqrt{-h}\left(
\bar\epsilon \Gamma^{\mu'\nu'}\Gamma^n D^+_{\mu'} \psi_{\nu'}
+ \bar \epsilon (K^{\mu'\nu'}-Kh^{\mu'\nu'})\Gamma_{\mu'}\psi_{\nu'}\right)\nn&&
-4\int_{\p\M} \delta_\epsilon A \wedge B\nn
\ea
Here we have used the identity
\be
 \Gamma^{\mu'\nu'} D_{\mu'}n_\lambda  \Gamma^\lambda
= K_{\mu'\lambda'}\Gamma^{\mu'\nu'}\Gamma^{\lambda'}
=(K^{\mu'\nu'}-Kh^{\mu'\nu'})\Gamma_{\mu'}
\ee
One can now see that these terms precisely cancel the terms in 
(\ref{newsuperJ}).

Note that we never had to deduce what $\delta_\epsilon \Phi_A$ was
from $\delta_\epsilon A$. However we have assumed that $A=d\Phi_A$
and $V=0$  on the boundary 
and these conditions impose constraints on the Fermions on the boundary (which
are certainly satisfied if the Fermions are on-shell).

Returning to the case of a general $A$ we see that if it is  non-trivial
on $\p\M$ then there is no boundary term that we can write down that
will cancel the Fermion boundary variation. Thus in these cases 
supersymmetry is broken by the presence of a  boundary.

\section{Higher-Dimensional Supergravity and an On-Shell 
Variant of New-Minimal}

Ultimately we are interested in extending our analysis to the ten and 
eleven-dimensional supergravities associated with string theory and
M-theory. As mentioned above, no
 off-shell completions are known for these theories and
it is generally believed that if any such formulations exist then they
must have an infinite number of auxiliary fields. 

However one can see that the type of physics explored here can be
extended in part to other supergravities. Suppose that there is a
supergravity Lagrangian ${\cal L}_{sugra}$ which is also invariant 
under a global symmetry (up to a boundary term).
Then we can make this symmetry local in the usual manner by replacing
covariant derivatives with gauge-covariant derivatives
\be
D_\mu \to {\cal D}_\mu = D_\mu + {\cal A}_\mu
\ee
where ${\cal A}_\mu$ is the appropriate gauge connection. In this way we obtain
the new Lagrangian
\be
{\cal L}_{{\cal A}} = {\cal L}_{sugra}(D_\mu \to {\cal D}_\mu)
\ee
Next we must arrange for ${\cal L}_{\cal A}$ to be supersymmetric. One sees 
that if we choose
\be
\delta_\epsilon {\cal A}_\mu  = 0 
\ee
then the variation of the action, ignoring boundary terms,  must be 
of the form
\be
\delta_\epsilon {\cal L}_{\cal A} = \Tr \left({\cal F}\wedge \Omega\right) 
\ee
where ${\cal F}$ is the gauge-invariant field strength of ${\cal A}$. 
To cancel such a term   we need only invent a new form field  ${\cal B}$ with  
\be
\delta_\epsilon {\cal B} =  \Omega
\ee
so that
\be
{\cal L}'_{\cal A}={\cal L}_{\cal A} - \Tr \left({\cal F}\wedge {\cal B}\right)
\ee
is supersymmetric, up to possible  boundary terms. 

In this way we have arrived a  form of supergravity that is similar
to  NMS. Of course the supersymmetry algebra is 
not closed off-shell, indeed we have merely added
a supersymmetry singlet $\cal A$, along with a non-singlet field ${\cal B}$. 
Thus these modifications may lead to problems in the quantum theory.

For example we could consider $N=1$ supergravity in four dimensions and
gauge the chiral $U(1)$ symmetry that new-minimal exploits. Following
the above procedure  we arrive at the Lagrangian
\be
\tilde {\cal L} = eR - 4ie\bar\psi_\mu\Gamma^{\mu\nu\rho}{\cal D}^+_\nu
\psi_{\rho} - 4F\wedge B
\label{newvar}
\ee
which is invariant  under
\begin{eqnarray}
\delta_\epsilon e_\mu^{\ \ \underline \nu} 
&=& -2i\bar\epsilon\Gamma^{\underline \nu}\psi_{\mu}\nonumber\\
\delta_\epsilon\psi_\mu &=& {\cal D}^+_{\mu}\epsilon \nonumber\\
\delta_\epsilon B_{\mu\nu} &=& 4i\bar\epsilon \Gamma_{[\mu}\psi_{\nu]}
\nonumber\\
\delta_\epsilon A_{\mu}&=& 0\nonumber\\
\end{eqnarray}
up to boundary terms. This is similar to, 
but not identical to, new-minimal with $V=0$. In addition it is
easy to see that  this theory can be made supersymmetric on a 
manifold with boundary using the same boundary terms and conditions
that we used for new-minimal.

\section{\sl Supersymmetric  Compactifications}

We can classify all supersymmetric 
compactifications of NMS 
supergravity (and also its variant (\ref{newvar})) 
of the form ${\bf R}^{4-d}\times {\cal M}_d$ with
$d=1,2,3$ and ${\cal M}_d$ compact without boundary. 
Our first condition on the manifold $ {\cal M}_d$, apart from compactness,  
is that it admits some kind of spinor and hence must be orientable
(we will not consider the possibility of pinors here).
The existence of a Killing spinor $\epsilon$ such that
${\cal D}^+_\mu\epsilon=0$ also implies that
\be
0=[{\cal D}^+_\mu,{\cal D}^+_\nu]\epsilon = 
\frac{1}{4}R_{\mu\nu\lambda\rho}\Gamma^{\lambda\rho}\epsilon
+F_{\mu\nu}\Gamma_5 \epsilon
\label{integrability}
\ee
Since $F=dA=0$ on-shell this 
implies that ${\cal M}_d$ is Ricci-flat and hence also Riemann flat
since $d=1,2,3$.

For $d=1,2$ the only possible internal manifolds are tori. 
These can clearly be made
supersymmetric. 
On the other hand we could also 
turn on a non-trivial $A = \frac{\alpha}{R_3}dx^3$
where $x^3$ is taken to be periodic with period $2\pi R_3$. 
The  Killing spinors take the form
\be
\epsilon = e^{-\alpha{x^3\over R_3}\Gamma_5}\epsilon_0
\ee
with $\epsilon_0$ a constant spinor. Only if $\alpha \in {\bf Z}$ 
do we find a single valued spinor. In this case the supercurrent 
will not be single valued so that the variation of the Lagrangian is
not exact and hence the   action not invariant. 
Thus a generic $A$ will apparently break all the supersymmetries.
However we can undo the damage  if we simply change the  
boundary conditions of the gravitino to 
$\psi_\mu(x^3+2\pi R_3)=e^{-2\pi\alpha\Gamma_5}\psi_\mu(x^3)$. 
In this case the supercurrent will be
single valued and hence the variation of the
Lagrangian is an exact form and the action invariant.

\subsection{\sl Compactification to one-dimension}

The case of $d=3$ is more interesting.
There are in fact  six compact orientable
Riemann flat three-manifolds called Bierberbach
manifolds (for example see \cite{wolf}).
They are all obtained as  quotients of ${\bf R}^3$ by some
freely acting group $G$ and
can be identified by their holonomies
\be
{\cal H}({\cal M}_3)= {\bf 1}\ ,\  {\bf Z}_2\ ,\   {\bf Z}_3\ ,\ 
{\bf Z}_4\ ,\   {\bf Z}_6\ , \  {\bf Z}_2 \times{\bf Z}_2
\ee

The first case is of course that of the torus 
${\bf T}^3={\bf R^3}/G$ with 
$G$ generated by the three elements
\ba
\alpha_1:(x^1,x^2,x^3) &\longrightarrow & (x^1+2\pi R_1,x^2,x^3)\nn
\alpha_2:(x^1,x^2,x^3) &\longrightarrow & (x^1,x^2+2\pi R_2,x^3)\nn
\alpha_3:(x^1,x^2,x^3) &\longrightarrow &(x^1,x^2,x^3+2\pi R_3)\nn
\ea
 
The first  Bieberbach manifold with nontrivial 
holonomy is a quotient  of ${\bf R}^3$  generated by
$\alpha^i$ along with the element $\beta$;
\be
\beta: (x^1,x^2,x^3) \longrightarrow (-x^1,-x^2,x^3+\pi R_3)\ .
\ee
so that $\beta^2 = \alpha_3$ and  $\beta\alpha_i\beta^{-1}=\alpha_i^{-1}$
if $i\ne 3$.
This leads to a space with holonomy ${\bf Z}_2$. (We have not written the 
most general such manifold: the lattice in the $12$ plane can be arbitrary
and need not be rectangular.)

The ${\bf Z}_4$ example is similar to the ${\bf Z}_2$ case only
now we take $R_1=R_2$ and  $\beta$ acts as
\be
\beta: (x^1,x^2,x^3) \longrightarrow (-x^2,x^1,x^3+{\pi\over 2} R_3)\ .
\ee
and hence we have $\beta^4 = \alpha_3$, $\beta\alpha_1\beta^{-1}=\alpha_2$
and $\beta\alpha_2\beta^{-1}=\alpha_1^{-1}$. 

Next we consider the $ {\bf Z}_3$ case. Here one must
fix $R_1=R_2$ and start with a hexagonal lattice, so that the
$\alpha_1$ generator is modified to 
\be
\alpha_1:(x^1, x^2, x^3) \to (x^1 + \sqrt{3}\pi R_1,x^2+\pi
R_1,x^3) 
\ee
while $\alpha_2, \alpha_3$ are unchanged. 
The generator $\beta$ is now
\be
\beta: (x^1,x^2,x^3) \to\left(-\frac{1}{2} x^1+\frac{\sqrt{3}}{2}x^2,-\frac{\sqrt{3}}{2} x^1-\frac{1}{2} x^2 , x^3
+ \frac{2\pi}{3}R_3\right )
\ee
which satisfies $\beta^3 = \alpha_3$, 
$\beta\alpha_1\beta^{-1}=\alpha^{-1}_2$
and $\beta\alpha_2\beta^{-1}=\alpha_1\alpha^{-1}_2$.

Next we consider the ${\bf Z}_6$ case. Again we must
fix $R_1=R_2$ and start with a hexagonal lattice, so that the
$\alpha_1$ generator is modified to 
\be
\alpha_1:(x^1, x^2, x^3) \to (x^1 + 
\sqrt{3}\pi R_1 ,x^2-\pi
R_1,x^3) 
\ee
The generator $\beta$ is now
\be
\beta: (x^1,x^2,x^3) \longrightarrow
\left( \frac{1}{2}x^1+\frac{\sqrt{3}}{2}
x^2,-\frac{\sqrt{3}}{2} x^1
+\frac{1}{2} x^2 , x^3
+ \frac{\pi}{3} R_3\right )\ .
\ee
and satisfies $\beta^6 = \alpha_3$, $\beta\alpha_1\beta^{-1}=\alpha_2^{-1}$
and $\beta\alpha_2\beta^{-1}=\alpha_1\alpha_2$.

The final case has holonomy ${\bf
Z}_2\times {\bf Z}_2$. In addition to the $\alpha_i$ (defined to
generate a rectangular lattice)  we introduce three additional generators
\begin{eqnarray}
\beta_1: (x^1,x^2,x^3) &\longrightarrow& (x^1+\pi R_1,-x^2+\pi R_2,-x^3)\nonumber\\
\beta_2: (x^1,x^2,x^3) &\longrightarrow& (-x^1+\pi R_1,x^2+\pi R_2,-x^3+\pi R_3)\nonumber\\
\beta_3: (x^1,x^2,x^3) &\longrightarrow& (-x^1,-x^2,x^3+\pi R_3)\nonumber\\
\end{eqnarray}
which satisfy $\beta_i^2 = \alpha_i$, 
$\beta_i\alpha_j\beta_i^{-1}=\alpha_j^{-1}$ if $i\ne j$ and 
$\beta_1\beta_2\beta_3 = \alpha_1$.

It has been shown in \cite{Pfaffle} that there are no Killing
spinors on a Bieberbach manifold with nontrivial holonomy.
To see this one notes that a Killing spinor on a Bieberbach manifold
will lift to a Killing spinor on the covering space ${\bf R}^3$.
However it must lift to a Killing spinor which is invariant
under the  group $G$. 

In order to proceed we need to define
a lift of the  group $G$ to a group 
$\tilde G\subset Spin(3) $ acting 
on the spinor bundle of ${\bf R}^3$, \ie\ 
for each generator $g$ of $G$ we must find
an element $\tilde g \in\tilde G$ such that 
$\pi (\tilde g)=g$ and which preserves the relations of
the group $G$. Here
$\pi:Spin(3)\to SO(3)$ is the usual 2-1 map. As detailed
in   \cite{Pfaffle} for each  group $G$ 
there will generically be several choices for $\tilde G$
and these correspond to different spin structures on the
Bieberbach manifold. 

Next we must ask that the Killing spinor is invariant. This
leads to a condition
\be
\tilde g \circ \epsilon(g\circ x) = \epsilon(x)
\label{spincon}
\ee
Since there is a unique spin bundle on the covering
space  we may choose a frame on ${\bf R}^3$ so that 
the Killing spinors are just  constant spinors $\epsilon=\epsilon_0$. 
The condition
(\ref{spincon}) is simply that 
$\tilde g \circ \epsilon_0= \epsilon_0$. 
We now note that all of the non-trivial Bieberbach manifolds
contain a generator $\beta$ which includes  a rotation in some plane 
by an amount different from  $2\pi$. The lift
of such a generator is an element $\tilde \beta\in Spin(3)$
such that $\tilde \beta \ne 1$. Hence it is impossible 
to find a constant spinor $\epsilon_0$ such that $\tilde \beta \circ
\epsilon_0 = \epsilon_0$.

However if we turn on the flat gauge connection
\be
A = \frac{1}{2R_3}dx^3
\ee 
then we can construct
invariant spinors. To see this note that the Killing spinors 
on ${\bf R}^3$, \ie\ spinors which satisfy ${\cal D}_\mu^+\epsilon=0$, 
are now
\be
\epsilon = e^{ -\frac{x^3}{2R_3}\Gamma_5}\epsilon_0
\ee
where $\epsilon_0$ is a constant spinor. The invariance condition
(\ref{spincon}) is now 
\be
e^{ \frac{x^3}{2R_3}\Gamma_5}\tilde g  e^{ -\frac{(g\circ x)^3}{2R_3}\Gamma_5} \epsilon_0 = \epsilon_0
\ee
For the first four non-trivial Bieberbach manifolds the only
non-trivial generator is $\beta$ which acts as 
\be
x^3 \to x^3 + \theta R^3\ ,\qquad \left(\matrix{x^1\cr x^2}\right)
\to  \left(\matrix{\cos\theta&\sin\theta\cr -\sin\theta& \cos\theta\cr}\right)\left(\matrix{x^1\cr x^2}\right)
\ee
where $\theta = \pi,2\pi/3,\pi/4,\pi/3$ for the holonomies
${\cal H} = {\bf Z}_2,{\bf Z}_3,{\bf Z}_4,{\bf Z}_6 $ respectively.
The corresponding lift to $Spin(3)$ of $\beta$ is
\be
\tilde \beta = \pm e^{{\theta\over 2}\Gamma_{12}}
\ee
where the choice of sign reflects a choice of spin structure on ${\cal M}_3$.
The invariance condition is now simply
\be
\pm e^{{\theta\over 2}\Gamma_{12}}e^{ -\frac{\theta}{2}\Gamma_5} \epsilon_0 = \epsilon_0
\ee
This can be solved by choosing the spin structure corresponding to the
plus sign and projecting onto constant spinors that satisfy
\be
\Gamma_{03}\epsilon_0 = \epsilon_0
\ee
For the ${\bf Z}_2$ case one can also find a Killing spinor with the spin structure
corresponding to the the minus sign by taking $\Gamma_{03}\epsilon_0 = -\epsilon_0$.

A key property of the first four non-trivial Bieberbach  
manifolds that enables these Killing spinors to exist 
is that the lift of $G$ to 
$\tilde G$ is $\tilde G =  U(1)\subset Spin(3)$.
This allows the holonomy of the spinor induced by each generator 
to be canceled by the phase shift induced 
by the $U(1)$ gauge connection $A$. For the final Bieberbach manifold,
with holonomy ${\bf Z}_2\times{\bf Z}_2$, $\tilde G$ is not
contained in a $U(1)$ subgroup of $Spin(3)$ \cite{Pfaffle} 
and hence the holonomies cannot be canceled. 
Thus there are no Killing spinors.

How did this work?
In ordinary supergravity the gravitinos are sections of
$T^\star ({\cal M})\otimes S ({\cal M})$, where $S ({\cal M})$ is a spinor 
bundle and $T^\star ({\cal M})$ is the cotangent bundle. For the
manifolds constructed above there are no Killing spinors, \ie\ covariantly constant sections of  $S ({\cal M})$. In the
NMS the gravitinos are sections of 
$T^\star ({\cal M})\otimes S ({\cal M})\otimes L({\cal M})$ where
$L({\cal M})$ is an additional flat line bundle. The point is that
there are covariantly constant sections of 
$S ({\cal M})\otimes L({\cal M})$. 

Note that one might try to make the Beiberbach manifolds supersymmetric
in old minimal supergravity  by changing the
boundary conditions to $\psi_\mu(x^3+\theta R_3)=\Gamma_5\psi_\mu(x^3)$
\footnote{We thank J. Maldacena for discussion on this point.}.
Such a boundary condition is  not compatible with the possible 
spin structures of spacetime but in principle this could be rectified 
by taking 
the Fermions to be sections of a line bundle associated to 
chiral rotations, as is the case in NMS, 
although without including a connection.
However this approach is problematic as the supersymmetry variations 
of the auxiliary fields in old minimal supergravity are not chirally 
covariant.

This is reminiscent of  
the situation with spin$^c$  structures. In these
cases there are manifolds, for example ${\bf CP}^2$, which
don't admit any spinors at all, let alone  covariantly constant ones.
However  they do admit sections of the spin bundle tensored 
with a complex line bundle; $S ({\cal M})\otimes L({\cal M})$ ({\it e.g.}
see \cite{HP}).
Indeed this situation can arise in string theory and M-theory \cite{DLP}.
Typically the complex line bundle is not flat and so cannot be a
solution of NMS, at least without coupling to additional
fields.  However in NMS it is possible to include
Spin$^c$ manifolds in the off-shell formulation of the theory by taking
the gauge field $F=dA$ to be  non-vanishing and (cohomologically) non-trivial.
In this sense ${\bf CP}^2$ is no more problematic  than
$S^4$, \ie\ the theory is defined for such manifolds but they
do not satisfy the equations of motion.

\section{Cylindrical Spacetimes and Singletons}

Gauge degrees of freedom are typically thought of as
unphysical. However this is not necessarily the case if spacetime has
a boundary. 
Quite generally, putting a gauge theory on a spacetime with a spatial
boundary can lead to physical gauge modes that live on the boundary.
This happens, for example, in three-dimensional Chern-Simons gauge 
theory and in the theory of the fractional quantum Hall effect, where 
the boundary degrees of freedom are known as ``edge states.'' 
In the context of supergravity one can find some discussion of 
these ``singleton modes'' in \cite{Witten:1998wy, MMS,Gukov:2004id, Belov:2004ht}.  
Since NMS (and the variant (\ref{newvar})) 
has additional, but auxiliary, gauge
degrees of freedom as compared with old minimal supergravity
it is possible that one could in principle distinguish between them
by considering spacetimes with a boundary. In some cases we may
therefore hope to see residual gauge degrees of freedom propagating 
on the boundary. A particular class of spacetimes with a boundary
are the so-called cylindrical spacetimes ${\cal M} = {\cal M}'\times
{\bf R}$, where ${\bf R}$ is the time dimension and $\p{\cal
  M}'=\Sigma$. Thus the boundary of spacetime is $\Sigma \times {\bf R}$.
Note that $\Sigma$ could have several disconnected pieces.

We wish to show that, in NMS, there is a consistent
choice of boundary conditions so that the theory on a boundary
contains additional physical modes that propagate on the boundary
due to the auxiliary fields. Hence one can, in principle, physically
distinguish between different off-shell formulations of supergravity
and, for example,  determine the existence or non-existence 
of a given set of auxiliary fields.

\subsection{An example}

First it is helpful to review the discussion of appendix A in \cite{MMS}.
Consider a Bosonic action 
\be
S = \int_{\M} dA\wedge B
\ee
where $\M = {\cal M}'\times {\bf R}$ with  $\p{\cal M}'=\Sigma$.

To exhibit the singleton modes
on the boundary we must carefully consider the boundary conditions to ensure a
well posed variational problem. 
Expanding in terms of the temporal and 
spatial components of  $A$ and $B$ this action is
\be
S = \int dt \int_{\M'} d'A'\wedge B_0+ A_0\wedge d'B'+\dot  A'\wedge B'
-\int dt\int_{\Sigma} A_0\wedge B'
\label{actis}
\ee
We can proceed in two ways. 
We could invoke the boundary condition 
$A_0=0$ on $\Sigma$. 
Alternatively
we could simply add an additional boundary term to the theory
\be
S_\p =  \int dt\int_{\Sigma} A_0\wedge B'
\ee
to cancel the existing boundary term in (\ref{actis}). No boundary 
condition is now required on $A_0$ or  $B_0$. Presumably these
two appoaches are equivalent and in either case the gauge symmetry
is broken on the boundary.

Continuing we can integrate 
out $A_0$ and $B_0$ since they are non-dynamical to find
\be
d'A'=0 \qquad d'B'=0
\label{eqis}
\ee
which we solve by
\be
A'=d'\Phi_A\qquad B'=d'\Phi_B
\label{solis}
\ee
where $\Phi_A$ and $\Phi_B$ are arbitrary. 
Substituting this back into the action leads to  
\ba
S &=& \int dt\int_{\M'}d' \dot \Phi_A\wedge d'\Phi_B \nn
&=&\int dt\int_{\Sigma} \dot \Phi_A\wedge d'\Phi_B\nn
\label{bdryactis}
\ea
Here we see the propagating singleton modes on the boundary. 

One can think of these singleton modes as arising from pure gauge modes
which violate the boundary condition $A_0=0$. To illustrate this point 
we note that in order to obtain a well-defined
boundary value problem we can also choose the boundary condition
\be
A = d\Phi_A\ ,\qquad dB=0
\ee
with $\Phi_A$ arbitrary,  \ie\ $A$ is exact and $B$ closed  on $\p\M$. 
Although it is important to note that such a boundary condition
removes  topologically non-trivial gauge configurations.

In this case no gauge symmetries are broken by the boundary.
Let us proceed as above and integrate over the bulk $A_0$ and $B_0$
fields. By this we mean that we split  $A_0=a_0+\tilde A_0$, where
$\tilde A_0$ vanishes on $\p\M$ and $a_0$ has support on $\p\M$,
and then integrate over $\tilde A_0$. In this way we find
\be
S= \int dt\int_{\Sigma} \dot \Phi_A\wedge d'\Phi_B
- a_0\wedge d'\Phi_B
\ee
Finally  we observe that the boundary conditions imply that 
$a_0 = \dot \Phi_A$ and hence the boundary action vanishes.
Thus, in this case, there are no boundary modes.

Finally we can consider what happens in the case where the theory
is not quite topological but includes a standard kinetic term for $B$
\be
S = \int_{\M} dA\wedge B + \frac{1}{2}dB\wedge \star dB
\label{acttwois}
\ee
It is not longer so simple to integrate out $B_0$. However
we can see that if we choose the boundary conditions which 
break the gauge symmetries then there will be massless gauge
modes that propagate along the boundary. These can also be thought of as
Goldstone modes for the global symmetry resulting from  gauge
transformations which do not vanish on the boundary.
For further details on singleton modes in such theories 
see \cite{Gukov:2004id, Belov:2004ht,Belov:2005ze}.

\subsection{Singletons in NMS}

NMS and its variant (\ref{newvar}) contain the same $dA\wedge B$
coupling that we have just discussed. (In NMS there is also
a kinetic term for $B$ which is absent in the latter case.)
Therefore we expect that if we choose gauge symmetry violating 
boundary conditions then singleton modes will propagate along the
boundary. 

We saw that by adding a suitable boundary term we could ensure that
both these actions were supersymmetric on a manifold with boundary
provided that we imposed the correct boundary conditions and terms. 
In   particular we required that 
$A=d\Phi_A$ and $V=0$ on $\p \M$. These boundary conditions
  restrict the topology of the connection $A$ but preserve
the gauge symmetry in the presence of the boundary. Thus there will not
be any singleton modes in this case.

In the more interesting case that we do not want to, or cannot, restrict
the gauge field $A$ to be exact then supersymmetry will be broken by
the boundary. Furthermore given the previous discussion we expect
to see singleton modes. In the case of NMS there is a kinetic
term for $B$, just as in the action (\ref{acttwois}). However from 
the discussion of (\ref{acttwois}) it is clear that there will be 
singleton modes from the gauge 
symmetry if we impose the boundary condition $A_0=0$ on $\p\M$. 

In the case of the variant theory  (\ref{newvar}) there
is no kinetic term for $B$ and we can be more explicit. In particular
we choose the boundary condition $A_0=0$.  
Proceeding as before we can integrate out $A_0$
which leads to the constaint
\be
d'B'-i\bar\psi_i\Gamma_j\psi_kdx^i\wedge dx^j\wedge dx^k=0
\label{B'const}
\ee
Next we integrate out $B_0$ to find
\be
d'A'=0
\ee
Thus we can set $A'=d'\Phi_A$.
Substituting all this back into the action we find
\be
S =\int dt \int_{\M'} \sqrt{-g}R-4i\bar\psi_i\Gamma^{i0k}D_0\psi_k 
-4i\bar\psi_\mu\Gamma^{\mu i\nu}{\cal D}_i^+\psi_\nu 
-4d'\dot \Phi_A \wedge B'
\ee
Note that $D_0$ appears instead of ${\cal D}_0^+$.
Next we integrate  
the last term by parts and use the constraint (\ref{B'const}) 
\be
S= \int dt \int_{\M'} \sqrt{-g}R
-4i\bar\psi_\mu\Gamma^{\mu\nu\lambda}
e^{-\Phi_A\Gamma_5}D_\nu (e^{\Phi_A\Gamma_5}\psi_\lambda )
-4\int dt
\int_{\Sigma } \dot \Phi_A \wedge B'
\ee
Lastly we perform the field redefinition 
$\psi_\mu = e^{-\Phi_A\Gamma_5}\Psi_\mu$
and  arrive at the familar on-shell supergravity but with an
additional boundary term
\ba
S = \int dt \int_{\M'} \sqrt{-g}R
-4i\bar\Psi_\mu\Gamma^{\mu\nu\lambda}D_\nu\Psi_\lambda -4\int dt
\int_{\Sigma } \dot \Phi_A \wedge B'
\ea
We must still solve for the contraint (\ref{B'const}), be
precise about the Fermionic boundary conditions and include any appropriate
boundary terms, such as the 
Gibbons-Hawking term. However regardless of
how we do this it is clear that we will always have 
$B' = d\Phi_B+...$ where $\Phi_B$ is a Bosonic boundary mode and 
the ellipsis denotes Fermionic terms. 
For example if we assume that $A_0=B_0=0$, 
$\Psi_\mu = D_\mu\eta $ is pure gauge
and $R_{\mu\nu\lambda\rho}=0$ on the boundary then we have a well posed
boundary value problem and we find that 
\be
B' = d'\Phi_B + i\bar\eta \Gamma_iD_j\eta dx^i\wedge dx^j
\ee 
on $\p\M$ so that the Fermionic gauge modes also propagate along the boundary.
Note that as a consequence of their topological origin the singleton modes
do not come with a factor of $\sqrt{-g}$  and hence do not contribute to
Einstein's equation.

\subsection{ Supersymmetry transformation of $\Phi_B$ }

It is helpful to  consider the on-shell supersymmetries of NMS. These are
\ba
\delta_\epsilon e_\mu^{\ \ \underline \nu} 
&=& -2i\bar\epsilon\Gamma^{\underline \nu}\psi_{\mu}\nonumber\\
\delta_\epsilon\psi_\mu &=& {\cal D}^+_{\mu}\epsilon   \nonumber\\
\delta_\epsilon B_{\mu\nu} &=& 4i\bar\epsilon \Gamma_{[\mu}\psi_{\nu]}
\nonumber\\
\delta_\epsilon A_{\mu}&=& 0\nonumber\\
\ea
and one  can check that they preserve the on-shell conditions
$dA=V=0$, as they should.
Note that from the condition $V=0$ we must have  
\be
dB - i\bar \psi_\nu \Gamma_\lambda \psi_\rho dx^\nu\wedge
dx^\lambda\wedge dx^\rho =0 
\ee
If $\psi_\mu = {\cal D}^+_\mu \eta$ is pure gauge then
\be
dB = id\left(\bar \eta \Gamma_\lambda {\cal D}_\rho^+\eta
dx^\lambda\wedge dx^\rho\right)
-i\bar \eta \Gamma_\lambda{\cal D}_\nu^+ {\cal D}_\rho^+\eta dx^\nu\wedge
dx^\lambda\wedge dx^\rho
\ee
The second term will vanish on-shell so that
\be
B = d\Phi_B + i\bar \eta \Gamma_\lambda {\cal D}_\rho^+\eta dx^\lambda\wedge dx^\rho
\ee
for an arbitrary one-form $\Phi_B$.
Under a supersymmetry generated by $\epsilon$ we clearly have that
\be
\delta_\epsilon \eta = \epsilon
\ee
Using the expression above for $\delta_\epsilon B_{\mu\nu}$ we see that
\ba
2i\bar \epsilon \Gamma_{ \mu}{\cal D}^+_{\nu}\eta\ dx^\mu\wedge dx^\nu
&=& d\delta_\epsilon\Phi_B + (i\bar \epsilon \Gamma_\lambda {\cal
  D}_\rho^+\eta
+i\bar \eta\Gamma_\lambda{\cal D}_\rho^+\epsilon )dx^\lambda\wedge dx^\rho \nn
&=&d\left( \delta_\epsilon\Phi_B +i\bar\epsilon\Gamma_\lambda \eta dx^\lambda\right)+
2i\bar \epsilon \Gamma_{ \mu}{\cal D}^+_{\nu}\eta\ dx^\mu\wedge dx^\nu
\nn
\ea
Thus
\be
\delta_\epsilon \Phi_B =  -i\bar\epsilon \Gamma_{\mu} \eta dx^\mu
\ee
Hence we see that the gauge zero modes $\eta$ and $\Phi_B$ are related
by  supersymmetry.

If we are on a manifold with boundary and use the boundary conditions 
(\ref{Fbc}) then the preserved supersymmetry is 
$\epsilon_+$ and we must set $\eta_-=0$ on $\p\M$, 
where the signs denote the eigenvalue of 
$\xi \Gamma^n e^{2\Phi_A\Gamma_5}$. In this case we see that
\be
\delta_\epsilon \Phi_{B} 
=  -i\bar\epsilon_+ \Gamma_{\mu'} \eta_+ dx^{\mu'}
\ee
Thus only the component of $\Phi_B$ that is tangential to the boundary 
is related to $\eta_+$ by supersymmetry.

\section{\sl Discussion }

In this paper we have discussed various  aspects of the auxiliary
fields that arise in new minimal supergravity (NMS). 
In particular we showed that
there are compact three-manifolds with well-defined Killing spinors which
are not well-defined in old minimal supergravity or simple off-shell
supergravity.  We also showed that, subject to
suitable boundary conditions, the auxiliary fields actually give rise 
to physical on-shell degrees of freedom that reside on the boundary 
of spacetime. Thus one can in principle 
distinguish between different off-shell forms of supergravity using
on-shell physics. We also demonstrated how half of the supersymmetry 
could be preserved in NMS on a manifold with boundary, 
provided the gauge field is trivial on the boundary. This suggests 
that there might be interesting applications to brane world senarios where
a topologically non-trivial auxiliary gauge field would lead to 
supersymmetry breaking.
Finally we would like to address some related issues.

It would be worthwhile extending the discussion of the present paper to 
other off-shell formulations of supergravity. Apart from old and new
minmal supergravity there is also 
the  so-called $\beta$FFC formulation \cite{GMOV}. 
It was  observed in \cite{GN} that the $\beta$FFC formulation
can be understood as the coupling of NMS supergravity to a
compensating chiral multiplet  whose Bosonic content is a complex
scalar (along with an auxiliary field). The logarithm of the absolute
value of the scalar field is identified with 
the dilaton $\phi$ whereas its phase  is eaten by the
two-form $B$ to produce a dynamical two-form which is dualized to the
axion $a$. 

Let us describe the $\beta$FFC formulation  in  more detail.
The complex scalar of the compensating chiral  multiplet is given a 
non-vanishing chiral weight. In
particular, under a chiral transformation, its phase $\varphi$ is shifted;  
$\varphi\to \varphi - \chi$ while its absolute value is invariant.  
The chiral covariant derivative of $\varphi$ is therefore 
\be
{\cal D}^+_\mu \varphi =\partial_\mu\varphi+ A_\mu
\ee 
Hence the kinetic term for $\varphi$ introduces a quadratic term for the
chiral gauge field  $A$ in the Lagrangian. 
The resulting $A$ equation of motion now algebraically determines
$A$ in terms of $B$ and $\varphi$ to be \cite{GN}
\be
A =-d \varphi-4 \star dB 
\ee 
(Recall that without coupling to the compensating  chiral multiplet the
$A$ equation of motion ensured that $B$ was non-dynamical: $dB=0$.)
Thus in the $\beta$FFC formulation there is still a chiral gauge field
that couples minimally to the Fermions, only now it is determined by
$B$ and $\varphi$. Note that the 
equation of motion for $B$ is $d (e^{2\phi} d\star B)=0$ and hence
it is possible to have $dA \ne 0$ on-shell. 

However we cannot make the Bieberbach manifolds supersymmetric as we did for
NMS since if $A= dx^3/2R$ then we must have $\varphi = x^3/2R$
or $dB\ne 0 $. The former case is forbidden as there are
couplings of $\varphi$ to the Fermions in the Lagrangian which require
that $\varphi$ be single valued. In the latter case one sees that
a non-zero $dB$ will lead to a non-vanishing  energy-momentum 
tensor so that the Bieberbach manifolds will no longer satisfy the
Einstein equations (although  
this raises the possibility of interesting new supersymmetric ``flux
compactifications'').
 
It is important to observe  that the chiral symmetry that NMS
supergravity gauges is anomalous. This has been shown \cite{GGS} to
lead to supersymmetry anomalies in the quantum theory. Happily
all is not lost as a Green-Schwarz anomaly cancelation for NMS
supergravity has been found in \cite{Baulieu:1986dy,CO} 
and one can show that the
Bieberbach manifolds remain supersymmetric.

There has been some  debate in the literature as to
whether or not old minimal, NMS or the $\beta$FFC  formulation 
results from four-dimensional string theory
\cite{CFG,Ovrut,S,deBoer:1996kt,GN} (see also 
\cite{Gates:2003qi,Alexander:2004xd} for related 
discussions on the appearance of new minimal and the $\beta$FFC formulations). 
However the main message of this paper has been to show that there can be
hidden on-shell physics in the auxiliary fields and these  remain
largely unknown in higher dimensions. Therefore
to  make further contact with string theory it is important 
to develop NMS and other off-shell formulations further.
In particular it is  not clear how to couple NMS to chiral multiplets with
a potential, as is needed in string theory. 
The problem is that   the superpotential
must transform under chiral symmetries. One way to achieve this might be
to postulate a chiral multiplet with a complex scalar $\phi_0$ that
shifts under the chiral symmetry
\be
\delta_\chi \phi_0 = -\chi
\ee
in addition to the other scalars $\phi_I$ which are chirally invariant.
Therefore  the covariant derivative acts on $\phi_0$ as
\be
{\cal D}_\mu^+\phi_0 = D_\mu \phi_0 + A_\mu\ 
\ee 
and is the ordinary derivative on $\phi_I$.
If this could be incorporated into NMS 
then one could attempt to include couplings to a superpotential of the form
\be
W = e^{2i\phi_0}\tilde W(\phi_I)
\ee
where $\tilde W$ depends holomorphically on $\phi_I$.
This suggestion 
is reminicient of the $\beta$FFC formulation, coupled to a superpotential. 
Therefore one expects
similar effects whereby $B$ eats the real part of $\phi_0$
and becomes the dynamical axion and the auxiliary gauge field $A_\mu$ is
algebraically determined in terms of the other fields.

Finally, let us return to our motivation of formulating $M$-theory on 
Spin$^{c}$ manifolds. Of course, we do not want to introduce a new 
propagating degree of freedom through the Spin$^{c}$ connection. In 
\cite{DLP} this degree of freedom is part of the $B$-field, but it is not evident how to 
implement such a relation in general. Another problem one must face is reconciling the 
Spin$^{c}$ structure with the standard reality properties of the gravitino. 
Finally, anomaly cancellation arguments would need to be modified. 
For example, the quantization of $G$-flux of \cite{Witten:1996md} now 
becomes $[G]_{DR} = \left( \frac{1}{4}\bar p_1 + \frac{1}{2}(\bar c_1)^2 \right) 
{\rm mod }\bar H^4(Y,Z) $, where the overline denotes reduction modulo torsion, 
and $c_1$ is the Chern class of the Spin$^{c}$  structure. 
(See \cite{Bouwknegt:2003vb} for related discussion.) Thus, finding 
such a generalization of $M$-theory - if it exists - seems quite challenging. 
The results of this paper make it clear that in such a search, one must first 
decide on some choice of off-shell formulation of the theory.

\section*{\sl Acknowledgements}

We would like to thank B. Acharya for discussions on Killing spinors
for Bieberbach manifolds. N.L. was supported by a PPARC Advanced Fellowship
and in part by the grant PPA/G/O/2002/00475. He would also like
to thank Rutgers University for its hospitality while some of
this work was completed. The work of   GM is
supported in part by DOE grant DE-FG02-96ER40949.


\begin{thebibliography}{10}

\bibitem{SW}
M.~F.~Sohnius and P.~C.~West,
``An Alternative Minimal Off-Shell Version Of N=1 Supergravity,''
Phys.\ Lett.\ B {\bf 105}, 353 (1981).
``The Tensor Calculus And Matter Coupling Of The Alternative Minimal Auxiliary
Field Formulation Of N=1 Supergravity,''
Nucl.\ Phys.\ B {\bf 198}, 493 (1982).
``The New Minimal Formulation Of N=1 Supergravity And Its Tensor Calculus,''
CERN-TH-3205
{\it Talks presented at Nuffield Workshop on Quantum Gravity and Supergravity, London, England, Aug 3-21, 1981}

\bibitem{Ferrara:1978em}
S.~Ferrara and P.~van Nieuwenhuizen,
``The Auxiliary Fields Of Supergravity,''
Phys.\ Lett.\ B {\bf 74}, 333 (1978).

\bibitem{Stelle:1978ye}
K.~S.~Stelle and P.~C.~West,
``Minimal Auxiliary Fields For Supergravity,''
Phys.\ Lett.\ B {\bf 74}, 330 (1978).




\bibitem{Kaku:1978ea}
M.~Kaku and P.~K.~Townsend,
``Poincare Supergravity As Broken Superconformal Gravity,''
Phys.\ Lett.\ B {\bf 76}, 54 (1978).



\bibitem{BaggerWess}
J. Bagger and J. Wess, {\it Supersymmetry and Supergravity}, Princeton Univ. Press 1992.

\bibitem{West:1990tg}
  P.~C.~West,
  {\it Introduction To Supersymmetry And Supergravity},
World Scientific, 1990.



\bibitem{deWit:1981ug}
  B.~de Wit, P.~van Nieuwenhuizen and A.~Van Proeyen,
  Phys.\ Lett.\ B {\bf 104}, 27 (1981).

\bibitem{deWit:1981fh}
  B.~de Wit and M.~Rocek,
  ``Improved Tensor Multiplets,''
  Phys.\ Lett.\ B {\bf 109}, 439 (1982).

\bibitem{Moss}
I.~G.~Moss,
``Boundary terms for supergravity and heterotic M-theory,''
arXiv:hep-th/0403106.

\bibitem{LM}
H.~C.~Luckock and I.~G.~Moss,
Class. Quant. Grav. {\bf 6}, 1993, (1989).


\bibitem{wolf}
J.~Wolf, ``Spaces of Constant Curvature'', Publish or Perish, 
1999.


\bibitem{Pfaffle}
F.~Pf\"affle, 
`` The Dirac Spectrum of Bieberbach Manifolds'',
J.\ Geo.\ Phys.\ {\bf 35}, 4, 367.




\bibitem{HP}
S.~W.~Hawking and C.~N.~Pope,
``Generalized Spin Structures In Quantum Gravity,''
Phys.\ Lett.\ B {\bf 73}, 42 (1978).


\bibitem{DLP}
M.~J.~Duff, H.~Lu and C.~N.~Pope,
``AdS(5) x S(5) untwisted,''
Nucl.\ Phys.\ B {\bf 532}, 181 (1998)
[arXiv:hep-th/9803061].




\bibitem{Witten:1998wy}
  E.~Witten,
  ``AdS/CFT correspondence and topological field theory,''
  JHEP {\bf 9812} (1998) 012
  [arXiv:hep-th/9812012].


\bibitem{MMS}
J.~M.~Maldacena, G.~W.~Moore and N.~Seiberg,
``D-brane charges in five-brane backgrounds,''
JHEP {\bf 0110}, 005 (2001)
[arXiv:hep-th/0108152].

\bibitem{Gukov:2004id}
  S.~Gukov, E.~Martinec, G.~W.~Moore and A.~Strominger,
  ``Chern-Simons gauge theory and the AdS(3)/CFT(2) correspondence,''
  arXiv:hep-th/0403225.

\bibitem{Belov:2004ht}
  D.~Belov and G.~W.~Moore,
  ``Conformal blocks for AdS(5) singletons,''
  arXiv:hep-th/0412167.

\bibitem{Belov:2005ze}
  D.~Belov and G.~W.~Moore,
  ``Classification of abelian spin Chern-Simons theories,''
  arXiv:hep-th/0505235.




\bibitem{GMOV}
S.~J.~J.~Gates, P.~Majumdar, R.~N.~Oerter and A.~E.~van de Ven,
``Superspace Geometry From D = 4, N=1 Heterotic Superstrings,''
Phys.\ Lett.\ B {\bf 214}, 26 (1988).



\bibitem{GN}
S.~J.~J.~Gates and H.~Nishino,
``Will the real 4-D, N=1 SG limit of superstring / M theory please stand
up?,''
Phys.\ Lett.\ B {\bf 492}, 178 (2000)
[arXiv:hep-th/0008206].


\bibitem{GGS}
S.~J.~Gates, M.~T.~Grisaru and W.~Siegel,
``Auxiliary Field Anomalies,''
Nucl.\ Phys.\ B {\bf 203}, 189 (1982).

\bibitem{Baulieu:1986dy}
 L.~Baulieu and M.~Bellon,
``Anomaly Cancellation Mechanism In N=1, D = 4 Supergravity And Distorted
Supergravity With Chern-Simons Forms,''
Phys.\ Lett.\ B {\bf 169}, 59 (1986).



\bibitem{CO}
G.~Lopes Cardoso and B.~A.~Ovrut,
``A Green-Schwarz Mechanism For D = 4, N=1 Supergravity Anomalies,''
Nucl.\ Phys.\ B {\bf 369}, 351 (1992).



\bibitem{CFG}
S.~Cecotti, S.~Ferrara and L.~Girardello,
``Gravitational Supervertices In N=1, 4-D Superstrings,''
Phys.\ Lett.\ B {\bf 198}, 336 (1987).

\bibitem{Ovrut}
B.~A.~Ovrut,
``Superstrings And The Auxiliary Field Structure Of Supergravity,''
Phys.\ Lett.\ B {\bf 205}, 455 (1988).

\bibitem{S}
W.~Siegel,
``Superstrings Give Old Minimal Supergravity,''
Phys.\ Lett.\ B {\bf 211}, 55 (1988).


\bibitem{deBoer:1996kt}
  J.~de Boer and K.~Skenderis,
  ``Covariant computation of the low energy effective action of the  heterotic
  superstring,''
  Nucl.\ Phys.\ B {\bf 481}, 129 (1996)
  [arXiv:hep-th/9608078].


\bibitem{Gates:2003qi}
  S.~J.~J.~Gates, W.~D.~.~Linch and J.~Phillips,
  JHEP {\bf 0502}, 036 (2005)
  [arXiv:hep-th/0311153].


\bibitem{Alexander:2004xd}
  S.~H.~S.~Alexander and S.~J.~J.~Gates,
  arXiv:hep-th/0409014.





\bibitem{Witten:1996md}
  E.~Witten,
  ``On flux quantization in M-theory and the effective action,''
  J.\ Geom.\ Phys.\  {\bf 22} (1997) 1
  [arXiv:hep-th/9609122].

\bibitem{Bouwknegt:2003vb}
  P.~Bouwknegt, J.~Evslin and V.~Mathai,
  ``T-duality: Topology change from H-flux,''
  Commun.\ Math.\ Phys.\  {\bf 249} (2004) 383
  [arXiv:hep-th/0306062].


\end{thebibliography}
\end{document}